\documentclass[acus]{JAC2003}
\usepackage{graphicx}

\begin{document}
\title{Compensation of the Crossing Angle with Crab Cavities at KEKB\vspace{-3 mm}}
%\thanks{Work supported by ...}}
\author{\begin{minipage}{16.8 cm}\begin{center} T. Abe, K. Akai, M. Akemoto, A. Akiyama, M. Arinaga, K. Ebihara, K. Egawa, A. Enomoto, J.~Flanagan, S. Fukuda
% ...
BINP, Novosibirsk, Russia\end{center}\end{minipage}}

\maketitle

\begin{abstract}
Crab cavities have been installed in the KEKB B--Factory rings to compensate the crossing angle at the collision point and thus increase luminosity. The beam operation with crab crossing has been done since February 2007. This is the first experience with such cavities in colliders or storage rings. The crab cavities have been working without serious issues. While higher specific luminosity than the geometrical gain has been achieved, further study is necessary and under way to reach the prediction of simulation.

%Although these crab cavities have been working without serious issues, the gain in the luminosity has not been much higher than the geometrical gain, and the reason is under the study.
\end{abstract}

\section{KEKB B--FACTORY}
KEKB B--Factory\cite{KEKB} has been operating since 1999 for the collision experiment mainly at the $\Upsilon$(4S) resonance. KEKB consists of the low energy positron ring (LER) at 3.5 GeV, the high energy electron ring (HER) at 8 GeV, and the injector linac. Two beams collide at the Belle detector. The machine parameters are listed in Table~\ref{params}. The highest luminosity, $1.72\times10^{34}{\rm cm}^{-2}{\rm s}^{-1}$, was recorded in Nov. 2006. The peak luminosity became higher than the design by 70\% mainly due to smaller $\beta_y^*$ (6 mm vs. 10 mm), horizontal betatron tune closer to a half integer (LER:0.505 / HER:0.511 vs. 0.52), and higher stored current in the HER (1.35 A vs. 1.1 A). The daily integrated luminosity is as twice high as the design due to the Continuous Injection Mode as well as acceleration of 2 bunches  per an rf pulse at the linac. The electron cloud in the LER, which was much more severe than thought in the design, has been mitigated up to 1.8 A with 3.5 bucket spacing by solenoid windings for 2,200 m.

\begin{table*}[bhtp]
\begin{minipage}{11.2 cm}
\caption{\small Machine parameters of KEKB at its best before crab crossing, comparing to the design. All luminosities are recorded values at the Belle detector\cite{belle}. The best data were recorded on Nov. 16, 2006, except for the integrated luminosities.}\label{params}
\end{minipage}
\begin{center}
{\small
\begin{tabular}{||l|c|c c|c c|c||}
\hline\hline
 & &\multicolumn{2}{c|}{\textbf{Best}} & 
	\multicolumn{2}{c|}{\textbf{Design}} & \\
 & & \multicolumn{2}{c|}{\textbf{LER~~~~HER}} & \multicolumn{2}{c|}{\textbf{LER~~~~HER}} & \\
\hline  
Circumference & $C$ & \multicolumn{4}{c|}{3014} & m\\
Beam Energy &$E$& 3.5 & 8 & 3.5 & 8 & GeV\\
Stored beam current&$I$ & 1.65 & 1.33 &   2.6 & 1.1 & A\\ 
Number of bunches / ring &$N_b$ &
\multicolumn{2}{c|}{1389} &
\multicolumn{2}{c|}{5000} & \\
Bunch current &$I_b$& 1.19 & 0.96 &    0.52 & 0.22 & mA\\
Bunch spacing &$s_b$& \multicolumn{2}{c|}{1.8--2.4} & \multicolumn{2}{c|}{0.6} & m\\
Horizontal Emittance &$\varepsilon_x$ & 18 & 24 &  18 & 18 & nm\\
Horizontal $\beta$ at IP & $\beta^*_x$ & 59 & 56 &  33 & 33 & cm\\
Vertical $\beta$ at IP &$\beta^*_y$ & 0.65 & 0.59 &   1.0 & 1.0 & cm \\
Horizontal size @ IP & $\sigma_x^*$ & 103 & 116 &   77 & 77 & $\mu$m\\ 
Vertical size @ IP &$\sigma_y^*$ & 1.9 & 1.9  & 1.9 & 1.9 & $\mu$m\\
Hor. Beam-beam parameter & $\xi_x$ &0.115 & 0.075 &   .039 & .039 & \\
Ver. Beam-beam parameter & $\xi_y$ & 0.101 & 0.056   & .052 & .052 & \\
Bunch length & $\sigma_z$ & 7 & 6 & 4 & 4 & mm\\
Horizontal rossing angle & $\theta_x$ & \multicolumn{4}{c|}{22} & mrad\\
Luminosity &$\cal{L}$ &\multicolumn{2}{c|}{17.12} &  
\multicolumn{2}{c|}{10} & /nb/s\\
$\int$Luminosity /  1, 7, 30 days &  & \multicolumn{2}{c|}{1.23, 7.82, 30.21}  & 
\multicolumn{2}{c|}{$\sim 0.6, -, - $} & /fb\\ 
$\int$Luminosity, total & & \multicolumn{2}{c|}{720}  &
\multicolumn{2}{c|}{100 for 3 years} & /fb\\
\hline\hline
\end{tabular}
~~
\begin{minipage}{5.3 cm}
\begin{center}
\caption{\small Typical parameters for the crab crossing.}
\begin{tabular}{||l|c|c|c||}
\hline\hline
 Ring & LER & HER & \\
 \hline  
$\theta_x$ & \multicolumn{2}{c|}{22} & mrad\\
 $\beta_x^*$ & 80 & 80 & cm\\
 $\beta_x^C$ &  73 & 162 & m \\
$\mu_x/2\pi$ & 0.505 & 0.511 &\\
$\psi_x^C/2\pi$ & $\sim 0.25$ & $\sim 0.25$ & \\
$V_C$ & 0.95 & 1.45 & V\\
$\omega_{\rm rf}/2\pi$ & \multicolumn{2}{c|}{509} & MHz\\
\hline\hline
\end{tabular}
\label{crabparams}
\end{center}
\end{minipage}
}
\end{center}
\vskip -5mm
\end{table*}

\section{CRAB CROSSING}
One of the main design features of KEKB is the horizontal crossing angle, 22 mrad, at the interaction point (IP). Although there are a lot of merits in the crossing angle scheme, the beam-beam performance may degrade. The design of KEKB predicted that the vertical beam-beam parameter $\xi_y$  is as high as 0.05 if betatron tunes are properly chosen, and actually KEKB has already achieved $\xi_y \approx 0.055$. Thus the beam-beam issues associated with the crossing angle was not critical if $\xi_y$ is lower than 0.05 or so.

% (1) Easier separation of the beams. No special bending magnets are necessary near the IP. (2) Minimum synchrotron radiation from the incoming beams. (3) Less hitting of the detector by particles which lose energy at the collision by Bremsstrahlung. (4) Simplified design near the IP to allow space to install the compensation anti-solenoid. 

\begin{figure}[hbtb]
\centering
\vskip -4 mm
\includegraphics[width=7 cm]{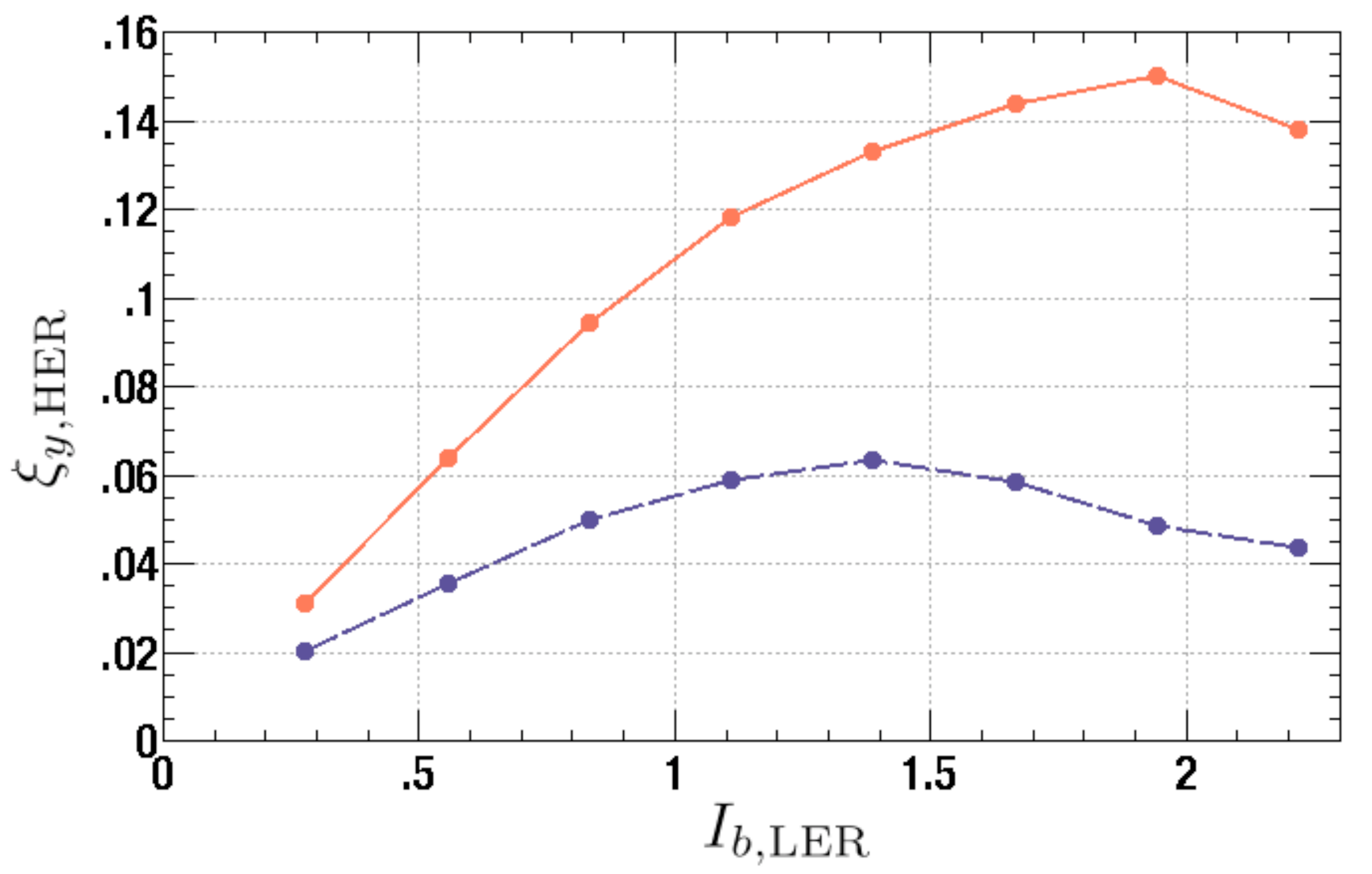}
\vskip -4 mm
\caption{\small Enhancement of the vertical beam-beam parameter by a head-on (crab) collision (upper curve) comparing to the crossing angle of 22~mrad (lower curve), obtained by a strong-strong beam-beam simulation.  Parameters are same as the present KEKB.}
\label{bbcrab}
\vskip -2 mm
\end{figure}

The crab crossing scheme, invented by R. Palmer\cite{Palmer}, was an idea to recover the head-on collision with the crossing angle. It has been also shown that the synchrotron-betatron coupling terms associated with the crossing angle are canceled by crab crossing\cite{OY}. The crab crossing scheme has been considered in the design of KEKB from the beginning as a backup solution for the crossing angle. Once, the crab crossing seemed non-urgent issue because KEKB achieved $xi_y \sim 0.055$ in the early stage of the operation (around the year 2000). Sooner or later an ineresting beam-beam simulation results appeared\cite{Ohmi}, predicting that the head-on or the crab crossing provides higher $\xi_y > 0.1$. Figure~\ref{bbcrab} shows the comparison of $\xi_y$ for the head-on (crab) and crossing angle with a strong-strong beam-beam simulation. Then the development of the crab cavities has been revitalized. 

\subsection{Single Crab Cavity Scheme}
\begin{figure}[bhtb]
\centering
\vskip -4 mm
\includegraphics[width=7 cm]{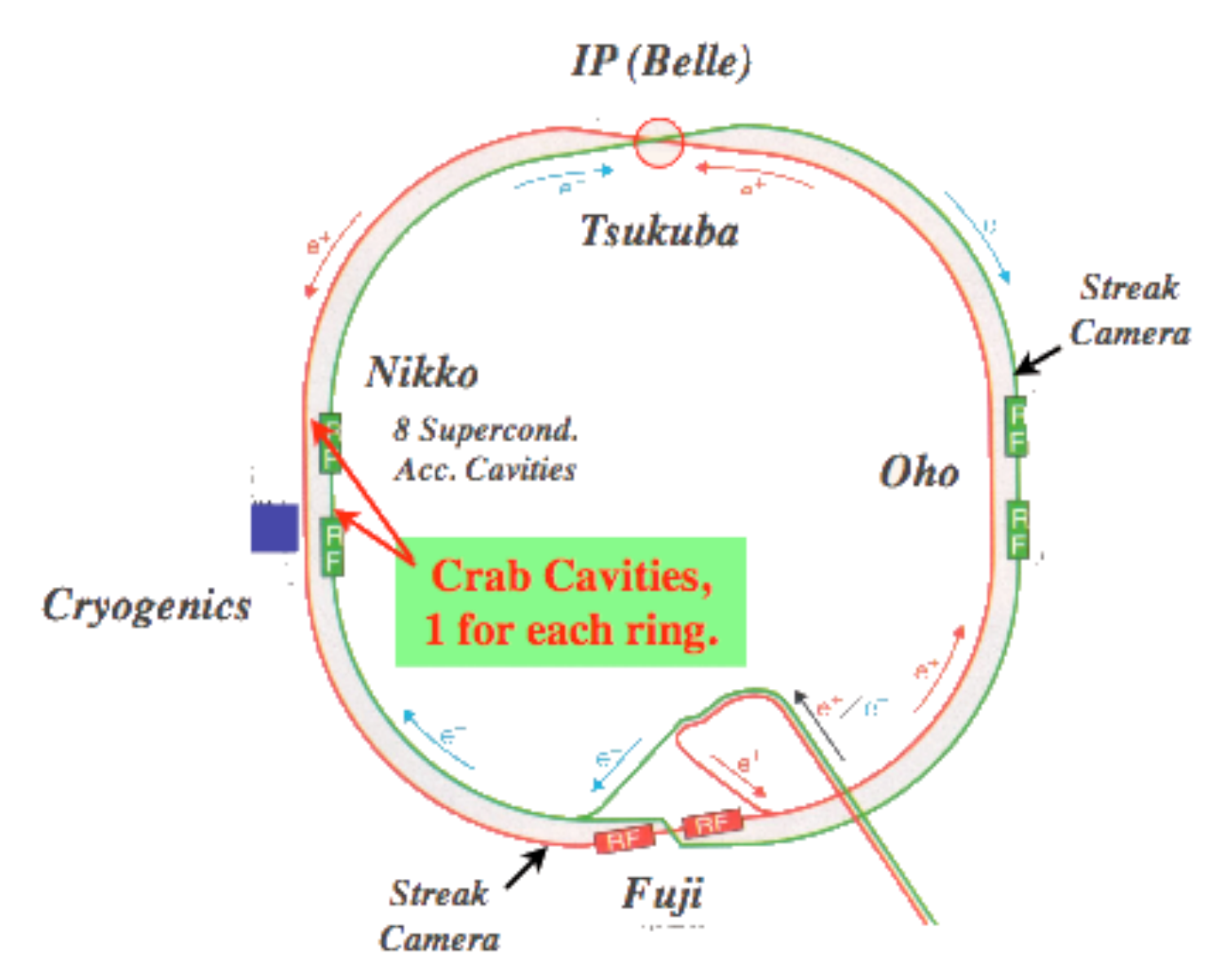}
\vskip -4 mm
\caption{\small The layout of the crab cavities at KEKB with the LER (red) and the HER (green). Each ring has one crab cavity at the Nikko straight section where the cryogenics already exists for the accelerating cavities. Each ring is equipped with a streak camera.}
\label{crablayout}
\vskip -2 mm
\end{figure}

The original design of KEKB had two cavities for each ring, both side of the IP, where the crab kick excited by the first cavity  is absorbed by another one. The new single crab cavity scheme extends the region with crab orbit until both cavities eventually merge to each other in a particular location in the ring. Then it needs only one cavity per ring. The layout is shown in Fig.~\ref{crablayout}. In the case of KEKB, this scheme not only saved the cost of the cavities, but made it possible to use the existing cryogenic system at Nikko for the superconducting accelerating cavities for the crab cavities. 

The beam optics was modified for the crab cavities to provide necessary magnitude of the $\beta$-functions at the cavities and the proper phase between the cavities and the IP\cite{amorita}. A number of quadrupoles have switched the polarity and became to have independent power supplies. The necessary horizontal kick voltage of the crab cavity $V_c$ must satisfy 
\begin{equation}
{\theta_x\over2}={\sqrt{\beta_x^C\beta_x^*}\cos(\psi_x^C-\mu_x/2)\over2\sin(\mu_x/2)}{V_C\omega_{\rm rf}\over Ec}\ ,\label{kick}
\end{equation}
where $\beta_x^C$, $\psi_x^C$, $\mu_x$, and $\omega_{\rm rf}$ are the horizontal $\beta$ function and the betatron phase relative to the IP at the crab cavity, tune of the ring, and the rf frequency of the crab cavity, respectively. The actual crab cavity
deflects the beam by a magnetic field, but we can define ``crab kick voltage" by the effective change in the transverse momentum. The resulting parameters for the crab cavities and beam optics are shown in Table~\ref{crabparams}. 
The rf frequency of the crav cavity is chosen equal to the accelerating cavities.

\section{CRAB CAVITIES}

\subsection{Design and Production}
\begin{figure}[htb]
\centering
\vskip -3 mm
\includegraphics[width=8 cm]{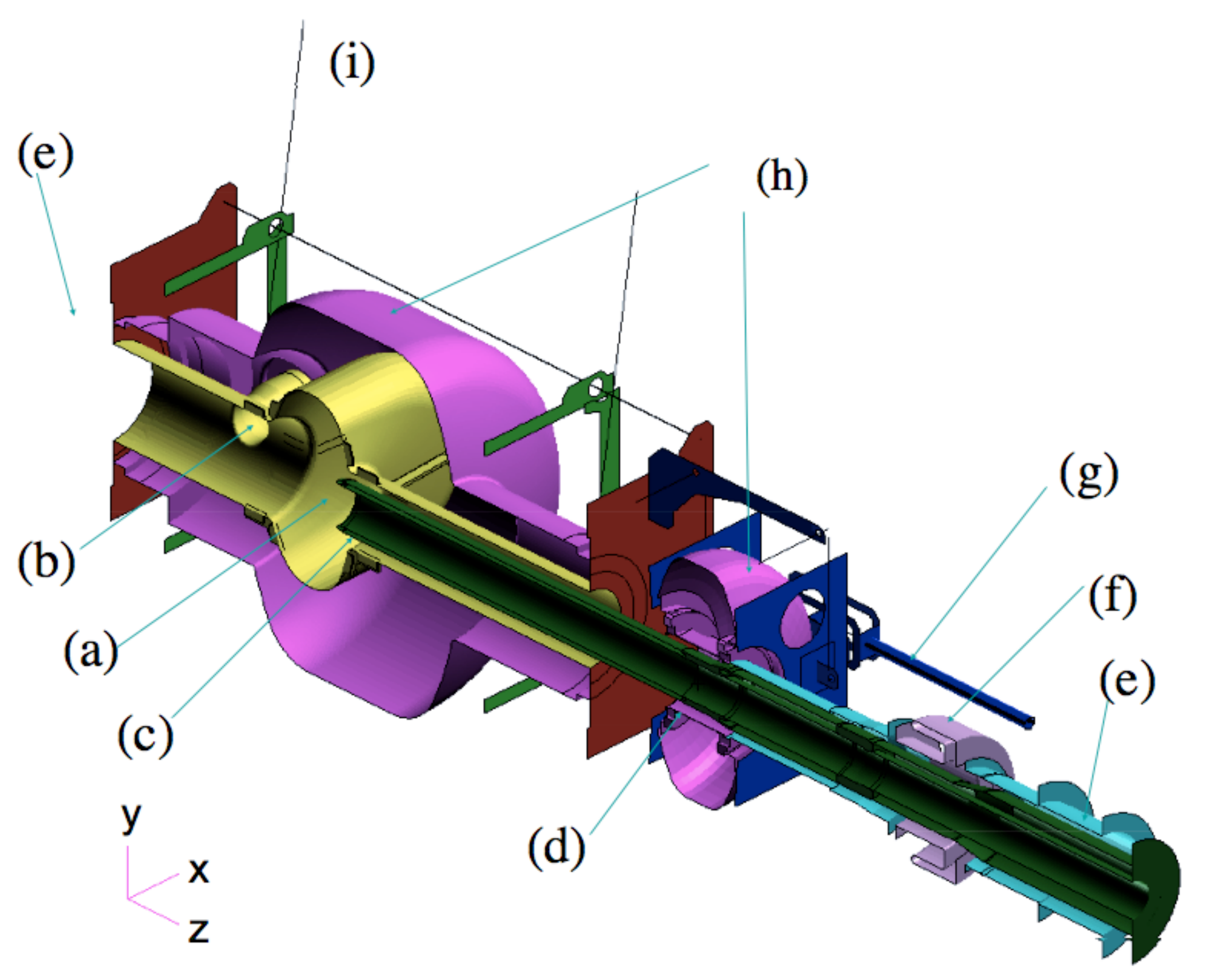}
\vskip -4 mm
\caption{\small KEKB crab cavity  consists of cavity cell (a), input coupler (b), coaxial beam pipe made of Nb (c) and its stub support (d), ferrite higher order mode (HOM) absorbers (e), notch filter (f) to reflect back the crab mode, tuning rod (g) of the coax, jacket type He vessels (h), and the support rods (i). The length in the z-direction shown in this figure is about 3 m, and the total system including the cryostat is about 6~m.}
\label{crabcav}
\vskip -2 mm
\end{figure}

The crab cavity for KEKB was originally designed by K.~Akai since 1991\cite{akai} and has been already included in {\it KEKB Design Report}. It is a superconducting cavity at 4K to use the lowest transverse mode to give the horizontal crab kick.  The main components of the crab cavity is shown in Fig.~\ref{crabcav}. The cavity is horizontally squashed so as to make the frequency of the vertical kick mode higher enough than the horizontal mode that is tuned to the accelerating rf frequency, 509~MHz. As it has an accelerating mode lower than the crab mode, a coaxial beam pipe is equipped to make it propagate out. The coax is also used as the frequency tuning of the crab mode by changing the insertion depth with a tuner rod externally driven by a piezo device and a pulsed motor located in room temperature. The higher order modes are basically damped by two absorbers made of ferrite, one in the large beam pipe and the other at the end of the coax. Some parasitic modes excited around the coax was mitigated by tilting the rod in the notch filter\cite{ymorita}. The engineering design of the crab cavity, cryostat, and periferal devices  was done by K.~Hosoyama and a number of prototypes have been tested since 1994 by his group\cite{hosoyama}, and finally converged into the present one. 

\subsection{Performance}
The  crab cavity for the HER was first assembled and tested in June 2006. Although the volage and Q-values satisfied the requirement, the cavity frequency were out of the tunable range. The reason was that the estimation of the relative thermal contraction between the cavity and the coax was not quite adequate. After the correction of the alignment, the second test was done in October 2006, then all performance satisfied the goal. The cavity for the LER was assembled and tested in December 2006, and the result was satisfactory. Both cavities were conditioned up to 1.8~MV kick voltage, with the unloaded Q--values higher than $10^9$ at the design voltage, 1.4~MV. The conditioning took less than a few days for the both cases.

\begin{figure}[htb]
\centering
\vskip -2 mm
\includegraphics[width=4 cm]{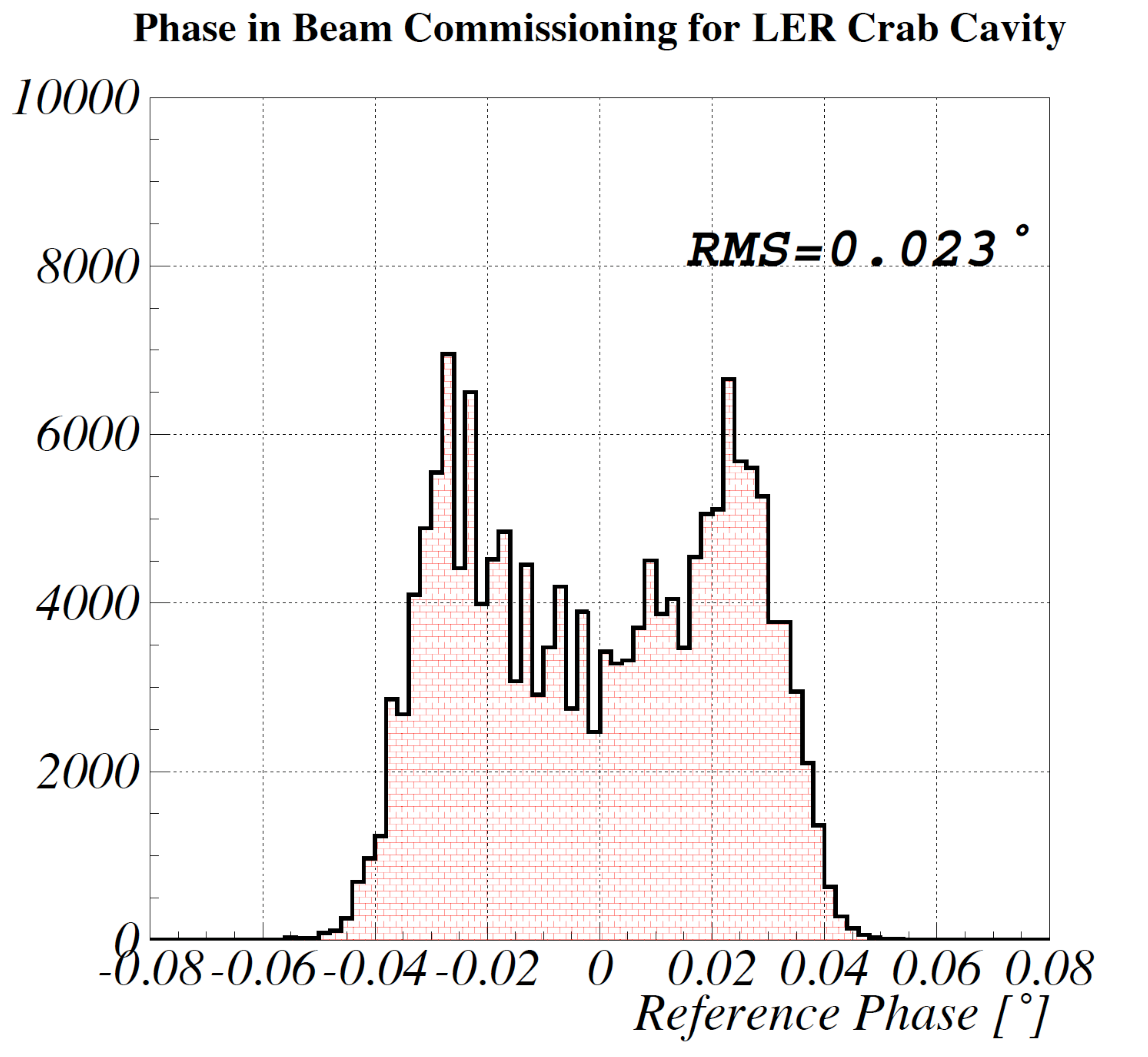}
\includegraphics[width=4 cm]{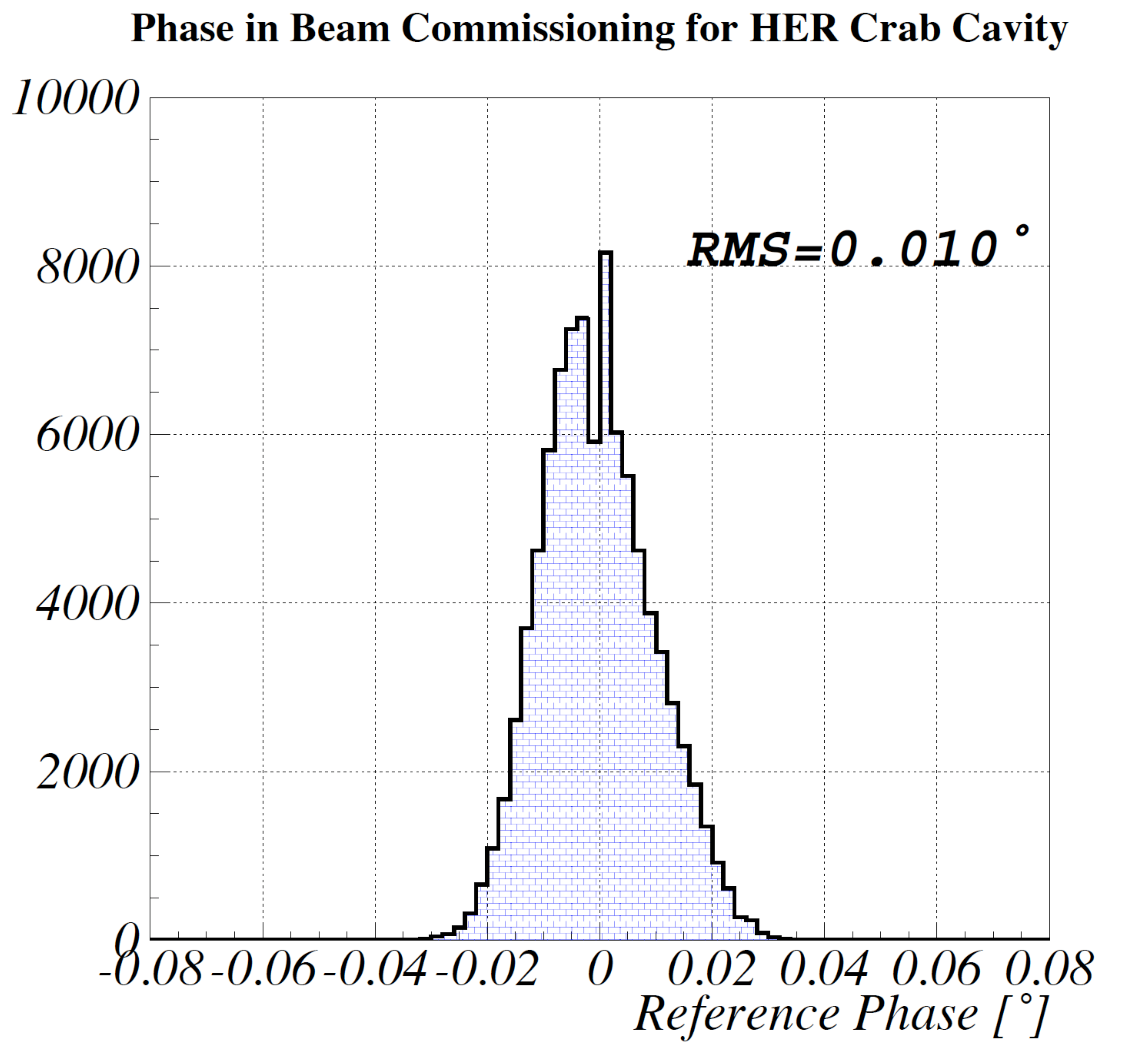}
\vskip -5 mm
\caption{\small Phase distribution of the crab cavities for the LER (left) and the HER (right) with the rf feedback. The standard deviations of the phases are 0.023 deg and 0.010 deg for the LER and the HER, respectively. }
\label{phase}
\vskip -2 mm
\end{figure}

As an example to show the performace of the crab cavities, Fig.~\ref{phase} shows the rf phase stability achieved with the rf feedback. The requirements for the phase fluctuation was achieved for both cavities. The reason why the LER has larger fluctuation than the HER was that the movement of the coax for the LER was not smooth enough, showing some friction or backlash, whose real cause has not yet been identified. The data in Fig.~\ref{phase} were taken by a data logger with a time constant of 1 second.  An independent measurement with a spectrum analyzer was done to detect faster phase fluctuation. The results of the phase fluctuation were within $\pm$0.01 degree for frequency region higher than 2 kHz, and within $\pm$0.1 degree between about 20 Hz and 2 kHz for the both rings. These results indicate that the high fluctuation in the LER in Fig.~\ref{phase} was in frequency region lower than 1 Hz. Both measurement  of the phase fluctuation with the data logger and the spectrum analyzer were well below the requirement. The measurement by the data logger  involves additional noise in phase detectors.

\begin{figure}[htb]
\centering
\vskip -3 mm
\includegraphics[width=4 cm]{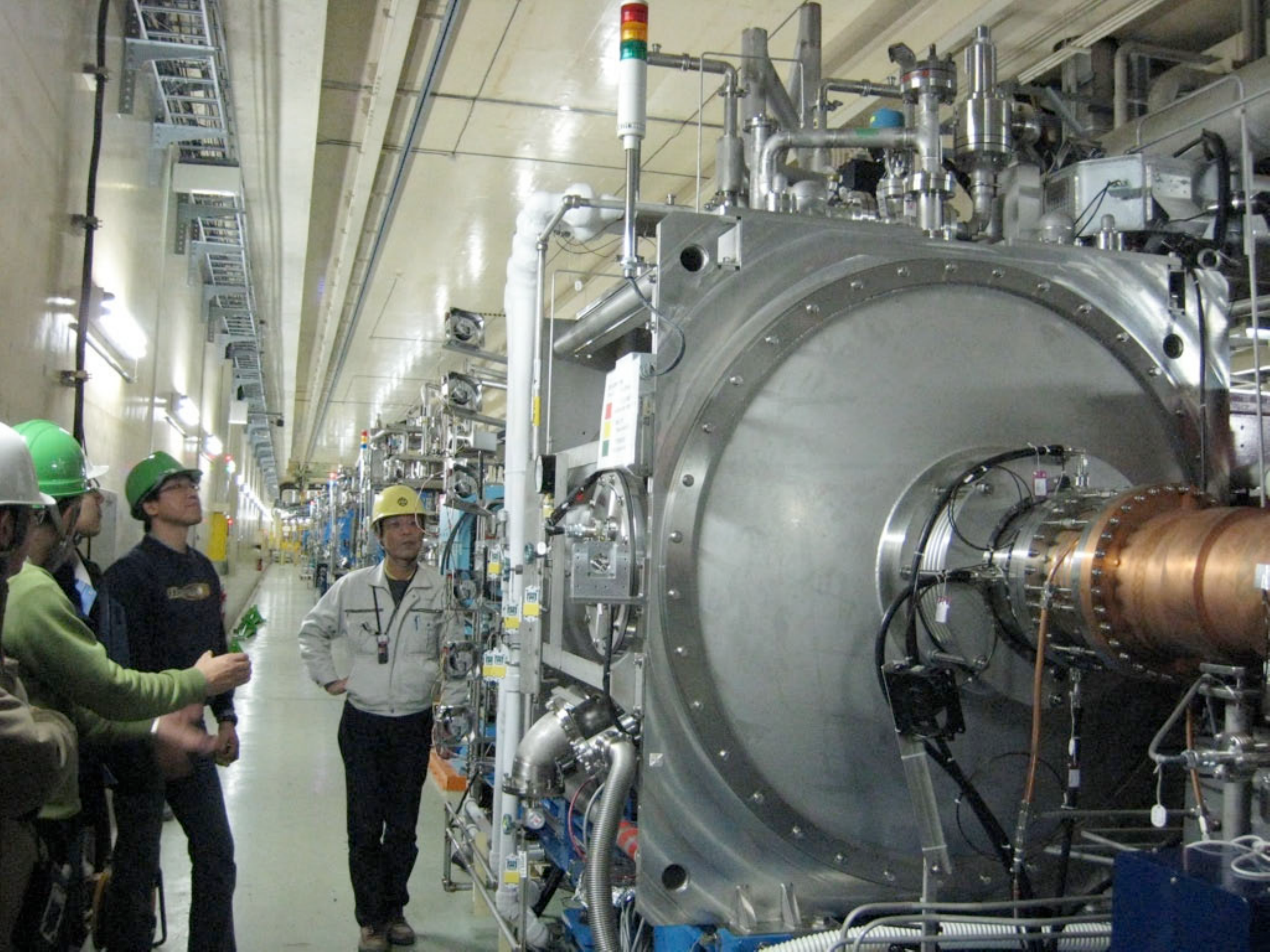}
\includegraphics[width=4 cm]{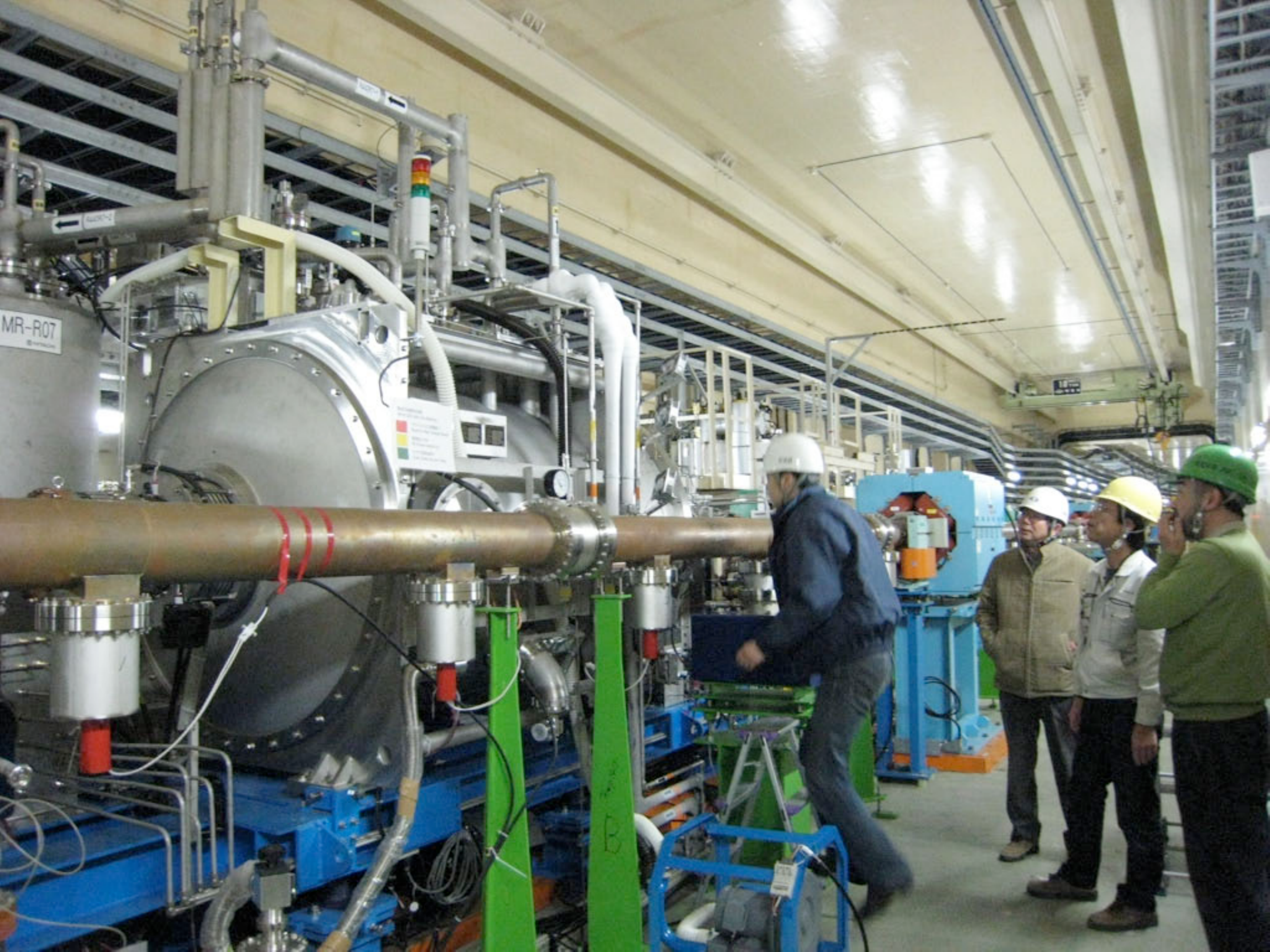}
\vskip -3 mm
\caption{\small Crab cavities were installed in the HER (left) and the LER (right), one for each ring.}
\label{cavsinring}
\end{figure}

The both cavities were installed in the rings in the winter shutdown from the end of December 2006 through January 2007, as shown in Fig.~\ref{cavsinring}. Further conditioning was done after the installation.  These cavities have been working with beam showing enough stability. These cavities have been warmed up three times to 80K, and once to 300K to remove the absorbed gas. The warm up to 300K significantlly reduced the rate of trips for both cavities, from 2 trips/day/cavity to 1 (HER) or 1/2 (LER) trips/day/cavity.

\section{FIRST BEAM TEST OF CRAB CROSSING}
The first beam test of the crab cavities started on February 14, 2007. After beam storage without crab voltage for a few days, the crab voltage was applied one by one. The tuning were mostly done with 50 or 100 bunches per ring in collision. The highest bunch current was kept below 1.5~mA in the LER, which was limited to protect the BPM electronics, and 0.5 to 0.7~mA in the HER.  The maximum luminosity with crab crossing was $10^{34}{\rm cm}^{-2}{\rm s}^{-1}$ with 1,000~mA and  540~mA in the LER and the HER, respectively, with 1091 bunches so far.

\subsection{Observation of Crabbing}
The very first test with the crab cavity was the observation of the kick in the horizontal orbit, changing the phase of the crab rf. It fits to a sine curve very well and the resulting kick voltages agree with estimation from the rf power in both rings within a few \% errors.

\begin{figure}[htbp]
\centering
\vskip -2 mm
\includegraphics[width=7 cm]{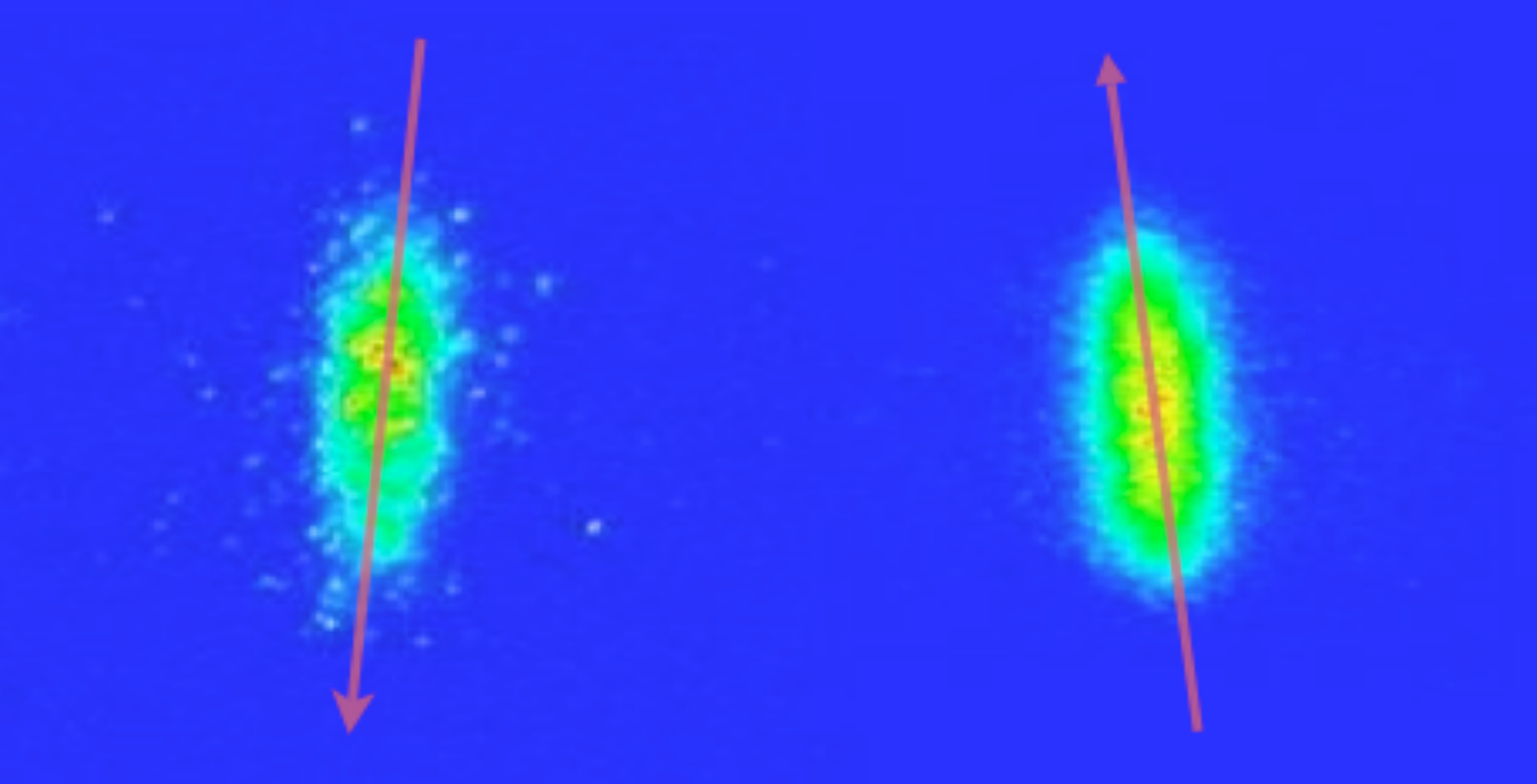}
\vskip -4 mm
\caption{\small Images taken by streak cameras, which locate as Fig.~\ref{crablayout},  show tilt of the bunches in the LER (left) and the HER (right)\cite{ikeda}.}
\vskip -2 mm
\label{streak}
\end{figure}

Then the tilts of the bunches were observed by streak cameras located in the rings. One of the merits of the single-cavity scheme is such an observation is possible, as the tilt is everywhere in the ring. Figure~\ref{streak} shows tilt of bunches. The response of the tilt to the phase of the crab cavities were right.

\subsection{Effective Head-on Collision}

As a result of the crab crossing, the response of the horizontal offset between two beams were greatly changed from with the crossing angle. With the crossing angle, the behavior of the beam, especially the vertical beam size of the LER was not symmetric to the sign of the offset\cite{funacol}. The vertical beam size blows up drastically when the HER beam comes inside at the IP relative to the LER. In this case the head of the LER beam collides to the HER beam with larger horizontal offset, as both beams comes from the inside to the outside at the IP. The longitudinal asymmetry of the LER bunch charge caused by the impedance, more charge at the head than the tail, is suspected as the cause of the asymmetry. Actually this asymmetry in the vertical beam size has been utilized to control the horizontal offset, as it has even better sensitivity than the horizontal beam-beam kick.

By the introduction of the crab crossing, such asymmetry disappeared or greatly reduced, since one of the sources of the asymmetry, the crossing angle,  went away. Then the feedback looking at the vertical size became unusable and feedback with horizontal beam-beam kick took place. This is a clear indication of the effective head-on collision.

\subsection{Specific Luminosity}
\begin{figure}[htb]
\centering
\includegraphics[width=8 cm]{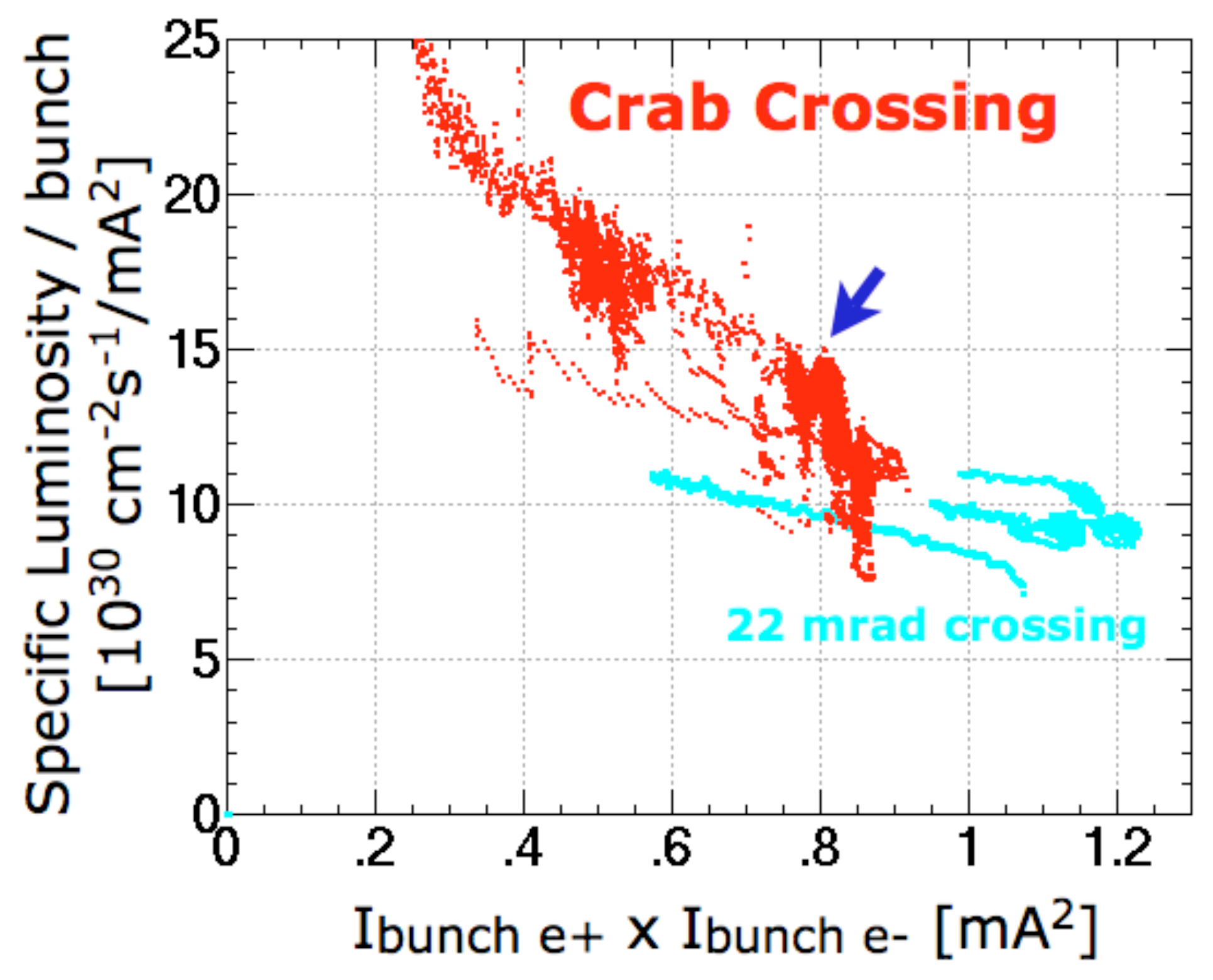}
\vskip -4 mm
\caption{\small Specific luminosity per bunch with crab crossing (red) comparing to crossing angle (blue) as a function of the product of bunch current. The highest $\xi_y\sim 0.088$ was achieved at the arrow.}
\label{speclum}
\vskip -3 mm
\end{figure}

Figure~\ref{speclum} shows the achieved specific luminosity per bunch as a function of the product of bunch currents, comparing the crab crossing and the crossing angle. The highest beam-beam parameter achieved with the crab crossing was 0.088, while it was 0.055 with crossing angle. The slope for the crab crossing data roughly follows a curve with $\xi_y\approx$constant, for the current product between 0.4 to 0.8 mA$^2$. The curve may seem steeper below 0.4, but the reason is not known. The product of the bunch did not become higher than 0.9, as the lifetimes dropped rapidly. The gain in the specific luminosity is higher than the geometrical one ($\sim 15\%$), but it has not reached the prediction of the beam-beam simulation.
\subsection{Tuning Parameters}

There are a number of knobs to tune up the crab crossing. Only a few of them can be tuned up with independent observables besides the luminosity. Table~\ref{knobs} lists the tuning parameters and its observables.

\begin{table*}[bhtp]
\begin{center}
\caption{Tuning knobs for the crab crossing and their observables. Many depend only on the beam size $\sigma_y$ at the synchrotron radiation monitor (SRM), besides the luminosity ${\cal L}$.}
{\small
\begin{tabular}{|| l | l | c ||}
\hline\hline
 Knob & Observable & frequency: every \\
 \hline  
Relative beam offset \@ IP & Beam-beam kick measured by BPMs around the IP & 1 sec\\
Relative beam angle \@ IP & BPMs around the IP & 1 sec\\
Global closed orbit & All $\sim$ 450 BPMs & 15 sec\\
Beam offset at crab cavities\cite{masuzawa} & BPMs around the crab cavity & 1 sec\\
Betatron tunes & tunes of non-colliding pilot bunches & $\sim$ 20 sec\\
Relative rf phase & center of gravity of the vertex & 10 min.\\
Global couplig, dispersion, beta-beat & orbit response to kicks \& rf frequency & $\sim$ 14 days\\
LER to HER crab voltage ratio & response in the hor. beam-beam kick. vs. crab rf phase& $\sim$ 7 days\\
Rf phase of crab cavity & hor. kick vs. crab voltage response & $\sim$ 7 days\\
Vertical waist position &  ${\cal L}$ and $\sigma_y$ at the SRM &  $\sim$1 day \\
Local x-y couplings and dispersions at IP & ${\cal L}$ and $\sigma_y$ at the SRM & $\sim$1 day each \\
Sextupole settings & ${\cal L}$ and lifetime & $\sim$ 3 days\\ 
X-y coupling parameter at the crab cavities & ${\cal L}$ and $\sigma_y$ at the SRM & $\sim$ 3 days\\
Crab kick voltage & ${\cal L}$ and $\sigma_y$ at the SRM & $\sim$ 7 days\\
\hline\hline
\end{tabular}
}
\label{knobs}
\end{center}
\vskip -7mm
\end{table*}

The issue is that so many knobs are optimized only by the luminosity and the beam size. The number of such knobs is about 30. It is in question that such multidimensional optimization actually reach the optimum starting with the large unknown errors. The tuning process of these knobs are slow due to the statistical error of the luminosity monitor especially at low current. Another slowing factor is that the data must be taken with the same beam currents to minimize the current dependence for each setting of the knobs. 

In many cases, the optimum setting of the knob is different for the luminosity maximum from the size minimum of the beam with the knob. Usually the size minima are pursued for a few days scanning all knobs for a few cycles, then switch to luminosity optimum. It is not clear that this kind of algorithm is adequate. 

One of the peculiar knobs for the crab crossing is to control the vertical crabbing at the IP. There are conceivable sources of the vertical crabbing such as x-y coupling at the IP and at the crab cavities, tilt of the accelerating cavities. The x-y coupling knobs at the IP and at the crab cavities can basically compensate such effects, but again there is no independent observable on this effect besides the luminosity and the beam sizes.

\subsection{Discussions}
The crab crossing has been tested at KEKB for about 4 months. The crab cavities has been working basically very well providing the necessary kick voltage stably. Although there are a lot of indications of the effective head-on collision, the specific luminosity has not reached the predicted value yet. There are a few speculation on the reason:
\begin{itemize}
\item Too many knobs are tuned only by the luminosity and the vertical beam size as described above.
\item The horizontal tunes are close to the synchrotron-betatron resonance line $2\nu_x+\nu_z={\rm integer}$. Actually single-beam  blowup of the horizontal and vertical beam sizes and drop of the beam lifetime were observed when the betatron tunes cross the resonance line.\cite{funa} The magnitude of the blowup strongly depend on the setting of the sextupoles. It is possible to estimate the blowup in the model by considering of the equilibrium horizontal emittance in the synchrotron phase space\cite{OK}. Such an optimization as well as the dynamic aperture has been tried to find out a good solution of sextupoles.
\item Negative momentum compaction factors have been tried in both rings to examine the effect of the resonance above, expecting a sum and difference resonances may behave differently. It was not successful, however, a longitudinal oscillation was found in the LER caused by a single bunch microwave instability.
\item There was a speculation related to the dynamic emittance effect caused by the beam-beam effect as the horizontal emittance largely increases when the horizontal tune is close to a half integer as KEKB (0.505 and 0.511). If the lattice has errors in the x--y coupling, such horizontal dynamic emittance may dilute to the vertical emittance. On the other hand, this effect can be cancelled if the local coupling at the IP is properly corrected.
\end{itemize}

\section{ACKNOWLEDGMENTS}
The authors thank H. Sugawara, Y. Totsuka, A. Suzuki, F. Takasaki, Y. Kamiya, S.-I. Kurokawa, M. Yamauchi, and the Belle Collaboration for supporting this program,  A. Hutton and the members of the KEKB Accelerator Review Committee for various suggestions for years, and R. Calaga and F. Zimmermann for useful discussions.

\end{document}